\documentclass[reprint,superscriptaddress,
amsmath,amssymb,
aps,
longbibliography,
pra, 
showpacs, 
floatfix,
pra,
]{revtex4-1}

\usepackage{graphicx}
\usepackage{array}
\usepackage{dcolumn}
\usepackage{bm}
\usepackage{color}
\usepackage{soul}

\newcolumntype{K}[1]{>{\centering\arraybackslash}p{#1}}

\makeatletter
\newcommand\xleftrightarrow[2][]{%
  \ext@arrow 9999{\longleftrightarrowfill@}{#1}{#2}}
\newcommand\longleftrightarrowfill@{%
  \arrowfill@\leftarrow\relbar\rightarrow}
\makeatother

\newcommand{\units}[2]{\ensuremath{#1\,\text{#2}}}
\newcommand{\ang}{\ensuremath{\text{\AA}}}
\newcommand{\expp}[1]{\ensuremath{\text{e}^{#1}}}

\newcommand{\omzero}{\ensuremath{\omega_{0}}}

\newcommand{\pol}{\ensuremath{\mathbf{P}}}
\newcommand{\magmom}{\ensuremath{\mathbf{M}}}
\newcommand{\que}{\ensuremath{\mathbf{Q}}}
\newcommand{\born}{\ensuremath{\mathbf{Z}^{\ast}}}

\begin{document}

\preprint{ }

\title{Dynamical Multiferroicity}

\author{Dominik~M.\ Juraschek}
\affiliation{Materials Theory, ETH Zurich, CH-8093 Z\"{u}rich, Switzerland}
\author{Michael\ Fechner}
\affiliation{Materials Theory, ETH Zurich, CH-8093 Z\"{u}rich, Switzerland}
\affiliation{Max Planck Institute for the Structure and Dynamics of Matter, DE-22761 Hamburg, Germany}
\author{Alexander~V.\ Balatsky}
\affiliation{Institute for Materials Science, Los Alamos, NM 87545, United States}
\affiliation{NORDITA, SE-106 91 Stockholm, Sweden}
\affiliation{Institute for Theoretical Studies, ETH Zurich, CH-8092 Z\"{u}rich, Switzerland}
\author{Nicola~A.\ Spaldin}
\affiliation{Materials Theory, ETH Zurich, CH-8093 Z\"{u}rich, Switzerland}

\date{\today}

\begin{abstract}

An appealing mechanism for inducing multiferroicity in materials is the generation of electric polarization by a 
spatially varying magnetization that is coupled to the lattice through the spin-orbit interaction. Here we describe 
the reciprocal effect, in which a time-dependent electric polarization induces magnetization even in materials with
no existing spin structure. We develop a formalism for this dynamical multiferroic effect in the case for which the 
polarization derives from optical phonons, and compute the strength of the phonon Zeeman effect, which is the 
solid-state equivalent of the well-established vibrational Zeeman effect in molecules, using density functional 
theory. We further show that a recently observed behavior -- the resonant excitation of a magnon by optically driven 
phonons -- is described by the formalism. Finally, we discuss examples of scenarios that are not driven by lattice 
dynamics and interpret the excitation of Dzyaloshinskii-Moriya-type electromagnons and the inverse Faraday effect from 
the viewpoint of dynamical multiferroicity.

\end{abstract}
                
\maketitle


\section{\label{Introduction} Introduction}

Multiferroic materials, with their simultaneous magnetism and ferroelectricity, are of considerable fundamental and 
technological interest because of their cross-coupled magnetic and electric external fields and internal order 
parameters. Particularly intriguing are the class of multiferroics in which a non-centrosymmetric ordering of the 
magnetic moments induces directly a ferroelectric polarization. Prototypical examples are chromium chrysoberyl, 
Cr$_2$BeO$_4$ \cite{Newnham_et_al:1978}, in which the effect was first proposed, and terbium manganite, TbMnO$_3$, 
which provided the first modern example and detailed characterization \cite{Kimura_et_al_Nature:2003,kenzelmann:2005}. 
The mechanism responsible for the induced ferroelectric polarization in these materials is the Dzyaloshinskii-Moriya 
interaction \cite{Dzyaloshinskii:1957,Dzyaloshinskii:1958,Moriya1:1960,Moriya:1960} as first proposed by Katsura 
\textit{et al.} \cite{katsura:2005}: Electric polarization, 
$\mathbf{P} \sim \mathbf{M} \times (\nabla_\mathbf{r} \times \mathbf{M})$, emerges in spatial spin textures as a 
consequence of the lowest order coupling between polarization and magnetization, 
$\mathbf{P}\cdot( \mathbf{M} \times ( \nabla_{\mathbf r} \times \mathbf{M}))$, in the usual Ginzburg-Landau picture 
\cite{katsura:2005,sergienko:2006,Mostovoy:2006,Cheong/Mostovoy:2007}.

From a symmetry point of view a reciprocal effect, in which magnetization is induced by ferroelectric polarization, 
must also exist. The time-reversal and spatial-inversion properties of magnetization and polarization -- 
$\mathbf{M}$ is symmetric under space inversion and anti-symmetric under time reversal, whereas $\mathbf{P}$ is 
symmetric under time reversal and anti-symmetric under space inversion -- indicate that the product 
$\mathbf{M} \cdot (\mathbf{P} \times \partial_t \mathbf{P})$ couples $\mathbf{M}$ and $\mathbf{P}$ at the same 
lowest order as $\mathbf{P}\cdot( \mathbf{M} \times ( \nabla_{\mathbf r} \times \mathbf{M}))$. It follows that a 
magnetization 
\begin{equation}\label{eq:dynmult}
\mathbf{M} \sim \mathbf{P} \times \partial_t \mathbf{P}
\end{equation}
develops in the presence of an appropriate dynamical polarization. This mechanism of induced magnetization by 
time-dependent polarization provides a dynamical analogy of the usual static multiferroicity. While the generation 
of polarization by spatial spin textures is now well established both experimentally and in first-principles 
calculations (see for example
Refs.~\cite{yamasaki:2007,malashevich:2008,kajimoto:2009,walker:2011}), discussions of the reciprocal dynamical 
effect are scarce, perhaps surprisingly so, since the idea is rooted in the classical induction of a magnetic field 
by a circulating current. One notable example is the proposal by Dzyaloshinskii and Mills \cite{dzyaloshinskii:2009} 
that it is the cause of the observed paramagnetism and specific heat increase in a ferroelectric insulator 
\cite{lashley:2007}; this analysis has been largely unrecognized to date. 

The purpose of this paper is to develop the formalism of this dynamical multiferroic effect and to describe 
observable effects that it leads to. We begin by deriving a formalism for magnetic moments induced by phonons that 
allows computation of their magnitudes using first-principles electronic structure calculations. This allows us to 
provide a general derivation of the phonon Zeeman effect as the solid-state equivalent to the well-established 
vibrational Zeeman effect in molecules and to analyze the limits under which the effect will be experimentally 
observable. Next, we give an explanation for a recently observed behaviour, the resonant magnon excitation of a 
magnon using optically driven phonons reported by Nova \textit{et al} \cite{nova:2017}, in terms of dynamical 
multiferroicity. Finally we discuss the connection of dynamical multiferroicity to effects that are not related to 
lattice dynamics, specifically Dzyaloshinskii-Moriya-type electromagnons and the inverse Faraday effect.

Our investigation is particularly timely in light of recent progress in the ability to manipulate materials in the 
time domain to generate nonequilibrium states of matter with properties that differ from or are inaccessible in the 
static limit. Examples include the ultrafast structural phase transition in (La,Ca)MnO$_3$ driven by melting of orbital 
order \cite{Beaud_et_al:2009} and the report of enhanced superconducting transition temperature in the high-$T_\text{c}$ 
cuprate yttrium barium copper oxide \cite{Hu_et_al:2014,mankowsky:2014,Kaiser_et_al:2014}. In addition, a number of 
theoretical proposals exist, for example predictions within Floquet theory of emerging topological states 
\cite{lindner:2011}, and of temporal control of spin-currents \cite{sato:2016}. Our work in turn contributes to this 
growing field of ``dynamical materials design'' by providing an additional mechanism -- time-dependence of polarization 
-- through which novel states can be dynamically generated.



\section*{Formalism for dynamical multiferroicity}

\begin{figure}[t]
\includegraphics[scale=0.16]{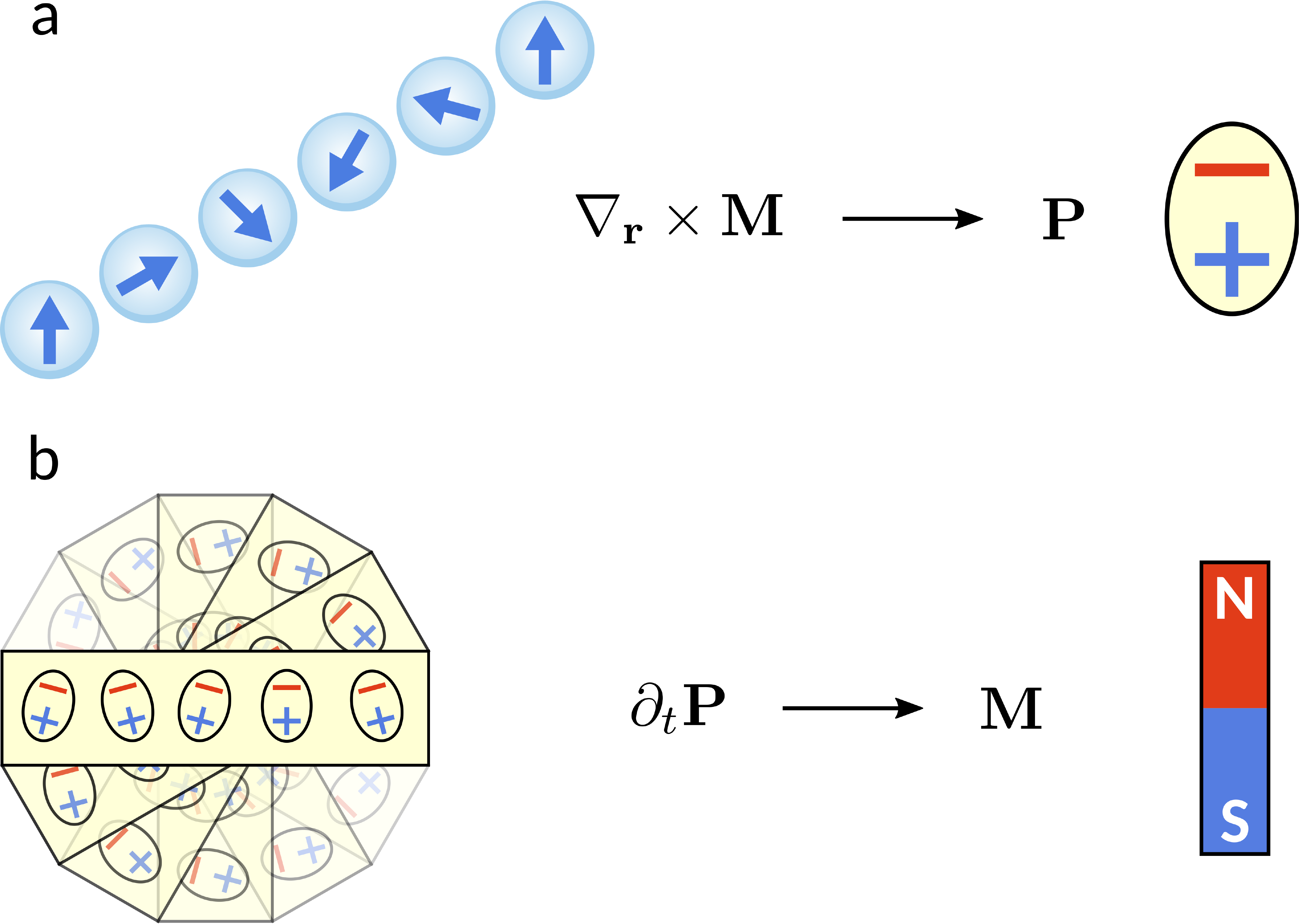}
\caption{
\label{fig:magpolduality}
\textbf{Duality of magnetization and polarization.}\\
\textbf{(a)} A spatially varying magnetization induces a polarization, as is well known for example in multiferroic
TbMnO$_3$ \cite{Kimura_et_al_Nature:2003,kenzelmann:2005,katsura:2005, Mostovoy:2006}. \textbf{(b)} A temporally varying 
polarization induces a magnetization as shown in this work.
}
\end{figure}

We begin by reviewing the duality between the time dependence of $\mathbf{P}$ and spatial gradient of $\mathbf{M}$ 
that we discussed in the introduction and that is illustrated in Fig.~\ref{fig:magpolduality}. A spatially varying 
magnetic structure induces a ferroelectric polarization; if the gradient is zero, the polarization vanishes. Both, 
the sense of the spin spiral and the direction of polarization, persist statically in one of two degenerate ground 
states: If the sense of the spin spiral is reversed, the polarization inverts. In the case of the dynamical 
multiferroic effect, 
a temporally varying polarization induces a magnetization; if the time derivative is zero, the magnetization vanishes.
Here, both, the sense of the rotating polarization (i.e. the curl of $\mathbf{P}\times\partial_t\mathbf{P}$) and the 
direction of the magnetization, persist \textit{dynamically} in one of two degenerate ground states of the dynamical 
system: If the sense of the rotating polarization is inverted, the magnetization reverses. A quasi-static state of the 
magnetization is realized when the rotation of the polarization is steady. The induced magnetization can then couple 
to lattice and magnetic degrees of freedom of the system, as we will show in various examples. We emphasize that 
magnetism can be induced by a time-varying polarization in a previously non-magnetic system, just as a spatially 
varying magnetization can induce a polarization in a system that is previously nonpolar. 

For the general case of a magnetization induced by two perpendicular dynamical polarizations that are oscillating 
sinusoidally with frequencies $\omega_1$ and $\omega_2$ and a relative phase shift of $\varphi$ we write the 
time-dependent polarization as
\begin{equation}\label{eq:polarization}
\mathbf{P}(t) = \left( \begin{array}{c}  P_1(t)  \\
                             P_2(t)  \\
                                      0
           \end{array}\right)
=
\left( \begin{array}{c}  A_1 \sin(\omega_1 t + \varphi)  \\
                             A_2 \sin(\omega_2 t)  \\
                                      0
           \end{array}
    \right),
\end{equation}
where one of the components can, in principle, also be static ($\omega_1$ or $\omega_2$ equal to zero). Evaluating 
Eq.~(\ref{eq:dynmult}) with this polarization we obtain
\begin{eqnarray}\label{eq:magneticmoment}
\mathbf{M}(t) & \sim & \bigg[  \frac{\omega_{+}}{2} \sin\big(\omega_{-}t+\varphi\big) \nonumber \\
      & & - \frac{\omega_{-}}{2}\sin\big(\omega_{+}t+\varphi\big) \bigg] A_1 A_2 \hat{z}.
\end{eqnarray}
A time-varying magnetization is induced that is oriented perpendicular to the spatial orientations of both 
polarizations. The magnetization consists of a superposition of a large-amplitude oscillation with the difference 
frequency, $\omega_{-}=\omega_1-\omega_2$, and a small-amplitude oscillation with the sum frequency, 
$\omega_{+}=\omega_1+\omega_2$. If the frequencies are equal, $\omega_1=\omega_2$, the induced magnetization is 
static.

\section*{Dynamical multiferroicity mediated by phonons}

\begin{figure}[t]
\includegraphics[scale=0.1]{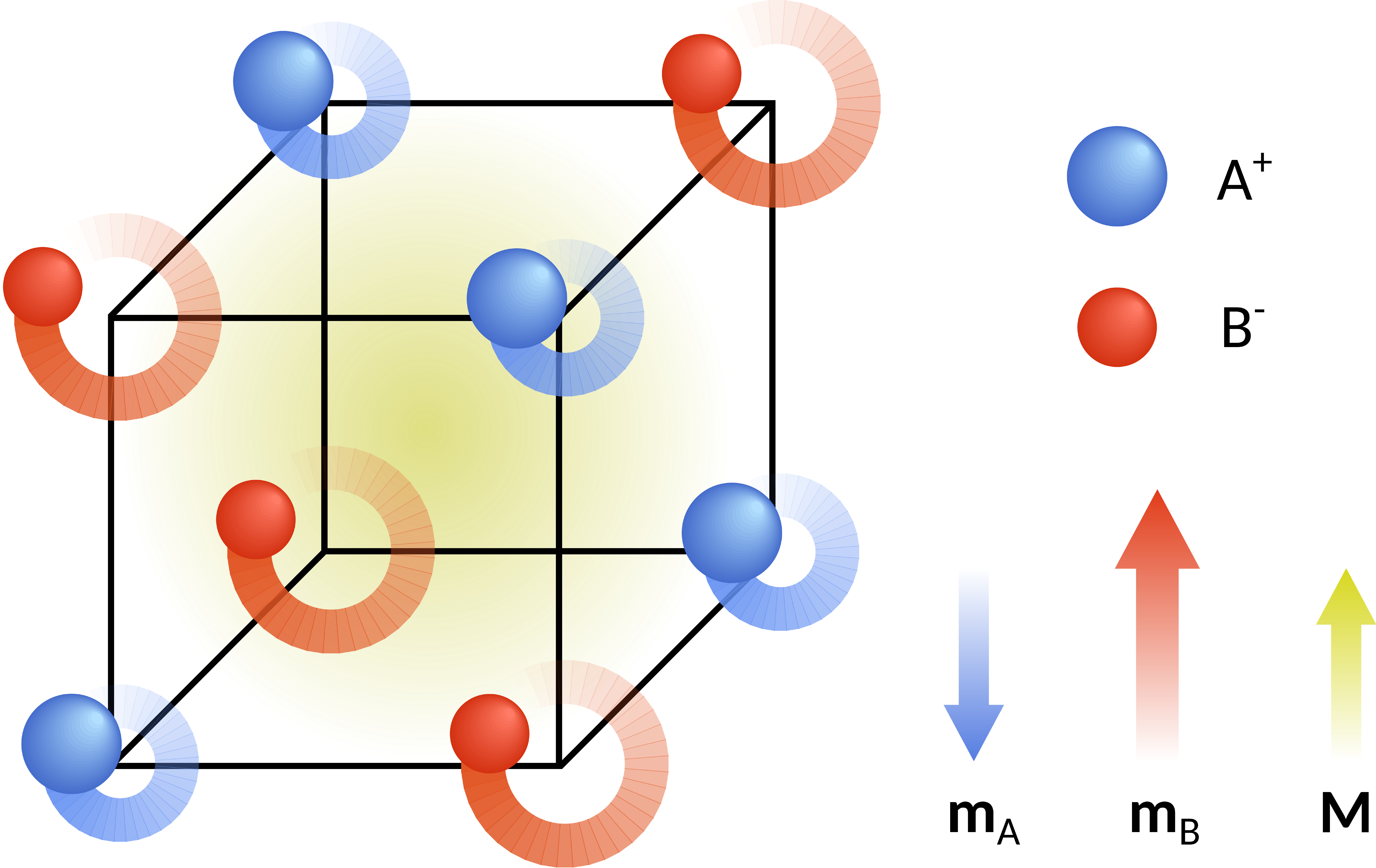}
\caption{\label{fig:phononmagneticmoment}
\textbf{Magnetic moments from ionic loops.}
Schematic motion of ions in a diatomic A$^{+}$B$^{-}$ material driven by perpendicular optical phonons. The circular 
motions of the ions create local magnetic moments, $\mathbf{m}_{\text{A}}$ and $\mathbf{m}_{\text{B}}$. The area covered
by the ionic loop of the lighter ion (here B$^-$) is larger, and therefore its local magnetic moment overcompensates 
that of the heavier ion. This leads to a macroscopic magnetic moment, \magmom{}.
}
\end{figure}

The polar nature  of optical phonons means that they can induce time-dependent polarization, and in turn
dynamical multiferroicity, in materials. We write the 
polarization, \pol{}, caused by the displacing atoms in a normal-mode vibration as the Born effective charges, 
\born{}, of the ions multiplied by the normal coordinates, \que{}, of the phonon modes, $\pol=\born\que$. 
(Note that another approach is to use a Berry phase calculation; this was shown to lead to similar results by 
Ceresoli and Tosatti \cite{ceresoli2:2002,ceresoli:2002}.) The 
connection to the usual relation between magnetic moment and angular momentum \cite{einsteindehaas:1915} is 
revealed by rewriting Eq.~\ref{eq:dynmult} to obtain 
\begin{equation}\label{eq:einsteindehaas}
\mathbf{M} = \gamma \mathbf{Q} \times \mathbf{\dot{Q}} = \gamma \mathbf{L},
\end{equation}
where $\gamma$ is a gyromagnetic ratio and $\mathbf{L}=\que\times\dot{\que}$ an angular momentum. Optical phonons 
with perpendicular polarity induce circular motions of the ions (ionic loops) whose individual magnetic moments combine 
to produce an effective macroscopic magnetic moment, as depicted in Fig.~\ref{fig:phononmagneticmoment}. The direction 
of $\mathbf{M}$ is determined by the sense of the ionic loop. The magnetic moment of phonons is the solid-state 
equivalent of the vibrational magnetic moment of molecules \cite{wick:1948}.

For a many-body system, such as a phonon, the simple tensorial relation of Eq.~(\ref{eq:einsteindehaas}) has generally 
to be extended. The total magnetic moment per unit cell given by the sum of the moments caused by the circular motion of 
each ion \cite{eshbach:1952}:
\begin{eqnarray}\label{eq:magmom_atomic}
\mathbf{M} = \sum\limits_{i} \mathbf{m}_{i} = \sum\limits_{i} \gamma_{i} \mathbf{L}_{i}.
\end{eqnarray}
Here $\mathbf{m}_{i}$ is the magnetic moment of ion $i$, $\mathbf{L}_{i}$ its angular momentum and $\gamma_{i}$ its 
gyromagnetic ratio, and the sum is over all ions in one unit cell. The angular momentum results from the motion of 
the ion along the eigenvectors of all contributing phonon modes \cite{huttner:1978}:
\begin{eqnarray}\label{eq:angular_atomic_detail}
\mathbf{L}_i  =  \sum\limits_{\alpha{},\beta{}} \mathbf{Q}_{i\alpha}\times\mathbf{\dot{Q}}_{i\beta}
                 =  \sum\limits_{\alpha{},\beta{}} Q_{\alpha}\dot{Q}_{\beta} \mathbf{q}_{i\alpha} \times 
                 \mathbf{q}_{i\beta},
\end{eqnarray}
where we wrote the displacement vector of ion $i$ corresponding to mode $\alpha$, $\mathbf{Q}_{i\alpha}$, in terms 
of a product of the normal mode coordinate amplitude, $Q_{\alpha}$, with the unit eigenvector, $\mathbf{q}_{i\alpha}$: 
$\mathbf{Q}_{i\alpha}=Q_{\alpha}\mathbf{q}_{i\alpha}$. Indices $\alpha{}$ and $\beta{}$ run over all contributing 
phonon modes. The gyromagnetic ratio tensor is the effective charge to mass ratio of the ion:
\begin{equation}\label{eq:gyro_atomic}
\gamma_{i}=\frac{e\mathbf{Z}^{\ast}_{i}}{2 M_{i}},
\end{equation}
where $\mathbf{Z}^{\ast}_{i}$ is the Born effective charge tensor of ion $i$ and $M_{i}$ its mass. 

We will now rewrite the magnetic moment of Eq.~(\ref{eq:magmom_atomic}) in terms of the phononic system. We insert 
the expression for the angular momentum of Eq.~(\ref{eq:angular_atomic_detail}) in Eq.~(\ref{eq:magmom_atomic}) to 
obtain: 
\begin{eqnarray}\label{eq:magmom_atomic_detail}
\mathbf{M} & = & \sum\limits_{i} \gamma_i
                \sum\limits_{\alpha{},\beta{}} Q_{\alpha}\dot{Q}_{\beta}
                \mathbf{q}_{i\alpha} \times \mathbf{q}_{i\beta} \nonumber\\
              & = & \sum\limits_{\alpha<\beta} 
              \big( Q_\alpha \dot{Q}_\beta - Q_{\beta} \dot{Q}_\alpha \big)
              \sum\limits_{i} \gamma_{i} \mathbf{q}_{i\alpha} \times \mathbf{q}_{i\beta}.
\end{eqnarray}
Now we write the difference of the normal mode coordinates as an angular momentum, analogously to 
Eq.~(\ref{eq:angular_atomic_detail}):
\begin{eqnarray}\label{eq:angular_phononic_detail}
\big( Q_\alpha \dot{Q}_\beta - Q_{\beta} \dot{Q}_\alpha \big) 
                   = \mathbf{Q}_{\alpha\beta} \times \dot{\mathbf{Q}}_{\alpha\beta} 
                   = \mathbf{L}_{\alpha\beta},
\end{eqnarray}
where $\mathbf{Q}_{\alpha\beta}$ contains the normal coordinates of the modes $\alpha$ and $\beta$ in the basis 
of their symmetric representation. The remaining part of \magmom{} resembles a gyromagnetic ratio, therefore we 
write Eq.~(\ref{eq:magmom_atomic_detail}) as: 
\begin{eqnarray}\label{eq:magmom_phononic}
\mathbf{M} = \sum\limits_{\alpha<\beta} \gamma_{\alpha\beta} \mathbf{L}_{\alpha\beta} = \sum\limits_{\alpha<\beta} 
\mathbf{m}_{\alpha\beta},
\end{eqnarray}
where 
\begin{eqnarray}\label{eq:gyro_phononic}
\gamma_{\alpha\beta} = \sum_{i} \gamma_{i} \mathbf{q}_{i\alpha} \times \mathbf{q}_{i\beta}
\end{eqnarray}
is the gyromagnetic ratio vector and $\mathbf{L}_{\alpha\beta}$ the angular momentum of a system of phonons. The 
induced magnetic moments, $\mathbf{m}_{\alpha\beta}$, are generated by the ionic loops caused by pairs of phonon 
modes, $\alpha$ and $\beta$. For only two contributing phonon modes Eq.~(\ref{eq:magmom_phononic}) reduces to the 
simple tensorial form
\begin{eqnarray}\label{eq:magmom_phononic_twomodes}
\mathbf{M} = \mathbf{m}_{12} = \gamma_{12} \mathbf{L}_{12} = \gamma_{12} \mathbf{Q}_{12} \times \dot{\mathbf{Q}}_{12}.
\end{eqnarray}
We note that all quantities in this section, particularly the Born effective charge tensors, $\mathbf{Z}^{\ast}_{i}$, 
and the phonon eigenvectors, $\mathbf{q}_{i\alpha}$, can be calculated from first principles using standard density 
functional theory methods.

\subsection*{Phonon Zeeman effect}

Next we discuss the Zeeman splitting of degenerate phonon modes, building on the early work extending 
the well-established vibrational Zeeman effect in molecules \cite{moss:1973} to solids by Anastassakis 
\cite{anastassakis:1972} and Rebane \cite{Rebane:1983}, as well as a derivation for cubic perovskites by Ceresoli 
\cite{ceresoli:2002} and a phenomenological analysis by Dzyaloshinskii and Mills \cite{dzyaloshinskii:2009}. 
Here we derive a general formalism for the phonon Zeeman effect that is amenable to computation using 
density functional theory via the quantities defined in the previous section.

Consider two degenerate phonon modes ($\omega_1=\omega_2=\omega_0$) polarized along perpendicular axes and shifted 
in phase by $\varphi \neq 0$. Eq.~(\ref{eq:magneticmoment}) then reduces to 
\begin{eqnarray}\label{eq:magmom_Zeeman}
\mathbf{M}(0) \sim \omega_0 \sin(\varphi) A_1 A_2 \hat{z}.
\end{eqnarray}
We see that the magnetization induced by the atomic motions is static and its magnitude 
depends only on the amplitude of the lattice vibrations, on their frequency, and on the phase shift between the two 
sinusoidal fields, reaching a maximum at $\varphi=\pi/2$. If an external magnetic field, \textbf{B}, is applied to 
the system, this induced magnetization interacts with it via the usual Zeeman coupling, with the result that the 
degeneracy of the phonon modes is lifted. 

\begin{figure}[t]
\includegraphics[scale=0.11]{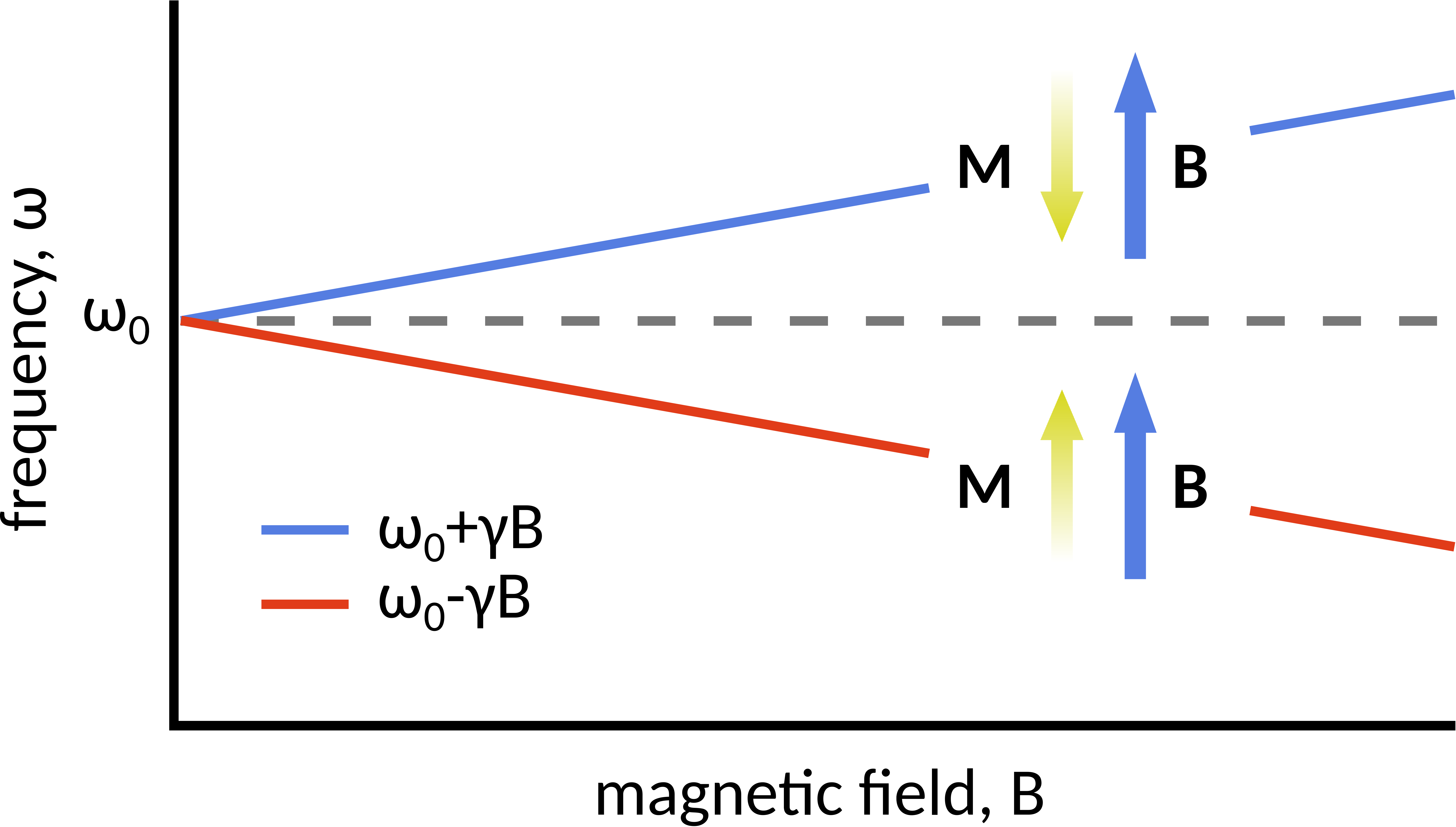}
\caption{\label{fig:phononzeemansplitting}
\textbf{Zeeman splitting of degenerate phonon modes.}
Sketch of the phonon Zeeman splitting as a function of the external magnetic field $\mathbf{B}$. The phonons with 
their magnetic moment $\mathbf{M}$ aligned parallel to the external magnetic field $\mathbf{B}$ has a lower energy 
than the phonons with their moment aligned antiparallel to the field.
}
\end{figure}

The Lagrangian describing the interaction of this magnetic moment with an externally applied magnetic field is:
\begin{eqnarray}\label{eq:lagrangianSUPP}
\mathcal{L}(\mathbf{Q},\mathbf{\dot{Q}}) = \frac{1}{2} |\mathbf{\dot{Q}}|^2
                                           - \frac{\omega^2}{2} |\mathbf{Q}|^2
                                           + \mathbf{B} \cdot \mathbf{M},
\end{eqnarray}
where $\mathbf{B}$ is the external magnetic field, $\mathbf{Q}=\mathbf{Q}_{12}=(Q_1,Q_2,0)$ contains the normal 
coordinates of two degenerate phonon modes and $\mathbf{M}$ is their magnetic moment as in 
Eq.~(\ref{eq:magmom_phononic_twomodes}). We can write the Lagrangian component wise as: 
\begin{eqnarray}\label{eq:lagrangian_detailed}
\mathcal{L}(Q_1,Q_2,\dot{Q}_1,\dot{Q}_2) & = & \frac{1}{2} \dot{Q}_{1}^{2} + \frac{1}{2} \dot{Q}_{2}^{2} -
                                             \frac{\omega^2}{2} Q_{1}^{2} - \frac{\omega^2}{2} Q_{2}^{2} \nonumber\\
                                         & & + \gamma B_z\big( Q_{1}\dot{Q}_{2} - Q_{2}\dot{Q}_{1} \big),
\end{eqnarray}
where $\gamma=\gamma_{12}$ is the gyromagnetic ratio as derived in 
Eqs.~(\ref{eq:gyro_phononic},\ref{eq:magmom_phononic_twomodes}), and $B_z$ is the $z$ component of the magnetic 
field. After a Fourier transformation,
\begin{align}
Q_{\alpha} & \rightarrow Q_{\alpha} = Q_{\alpha\omega}\expp{i\omega t} \\
Q_{\alpha}^2 & \rightarrow Q_{\alpha\omega}Q_{\alpha\omega}^{\ast} \\
Q_{\alpha}\dot{Q}_{\beta} & \rightarrow \frac{1}{2}\left(Q_{\alpha\omega}\dot{Q}_{\beta\omega}^{\ast}+
Q_{\alpha\omega}^{\ast}\dot{Q}_{\beta\omega}\right),
\end{align}
the Lagrangian becomes
\begin{equation}
\mathcal{L}(\mathbf{Q_{\omega}},\mathbf{Q_{\omega}^{\ast}}) = \mathbf{Q_{\omega}} \mathbf{A} 
\mathbf{Q_{\omega}^{\ast}},
\end{equation}
where $\mathbf{Q_{\omega}}=(Q_{1\omega},Q_{2\omega},0)$ contains the normal mode coordinates in Fourier space, and
\begin{equation}
\mathbf{A} = \left( \begin{array}{cc} \omega^2-\omega^2_0 & 2i\gamma B_z\omega \\ -2i\gamma B_z\omega & 
\omega^2-\omega^2_0 \end{array} \right).
\end{equation}
We solve the determinant for the zone centre ($\omega\rightarrow\omega_0$) and obtain a splitting of the degenerate 
phonon modes:
\begin{equation}\label{eq:phononZeemansplitting}
\omega = \omzero\sqrt{1\pm\frac{2\gamma B}{\omzero}} \approx \omzero \pm \gamma B_z .
\end{equation}
The splitting is sketched in Fig.~\ref{fig:phononzeemansplitting} for a magnetic moment aligned parallel and 
antiparallel to an external field, corresponding to a left- and right-handed sense of the phonons. Notably the phonon 
Zeeman effect is independent of the magnitude of the phonon magnetic moment and hence the amplitude of the lattice
vibrations. It is therefore present in optical modes excited, for example thermally or due to zero-point fluctuations 
\cite{dzyaloshinskii:2009,riseborough:2010}, and does not require intense optical pumping.

We estimate the magnitude of the effect for the tetragonal phase of strontium titanate, SrTiO$_3$, using density 
functional theory. (Please see the ``Computational methods for phonon calculations'' section for details of the 
calculation.) Our calculated low frequency degenerate E$_\text{g}$ phonon modes lie at \units{1}{THz}, and their 
gyromagnetic ratio as calculated with Eq.~(\ref{eq:gyro_phononic}) is $\gamma=0.9\times10^{7}\,\text{T}^{-1}\text{s}^{-1}$. 
(Note that the g-factor obtained by Ceresoli \cite{ceresoli2:2002} for cubic SrTiO$_3$ using the Berry phase approach has 
a similar value.) It follows that to obtain a relative splitting of the eigenfrequency of $2\gamma B/\omzero\approx 10^{-3}$ 
one needs a magnetic field of the order of $B=\units{55}{T}$; at this field strength, our splitting is four orders of 
magnitude larger than the value estimated by Dzyaloshinskii and Kats for the splitting of \textit{acoustic} phonons 
\cite{dzyaloshinskii:2011}. While the splitting for SrTiO$_3$ is small, the effect will be significantly enhanced in 
materials with high Born effective charges, light ions, and low-frequency optical phonon modes. Promising candidates are 
ABO$_3$ compounds in which a heavy A/B-site ion ensures low frequency phonons, while the light B/A-site ion and the oxygens 
make up a large percentage of the motion in the normal mode. A further approach to increasing the magnitude of the effect 
is to identify materials with dynamically varying Born effective charges which will lead to an additional contribution to 
the angular momentum proportional to $(\partial_t\mathbf{Z}^\ast)\mathbf{Q}$. 

\begin{figure*}[t]
\includegraphics[scale=0.27]{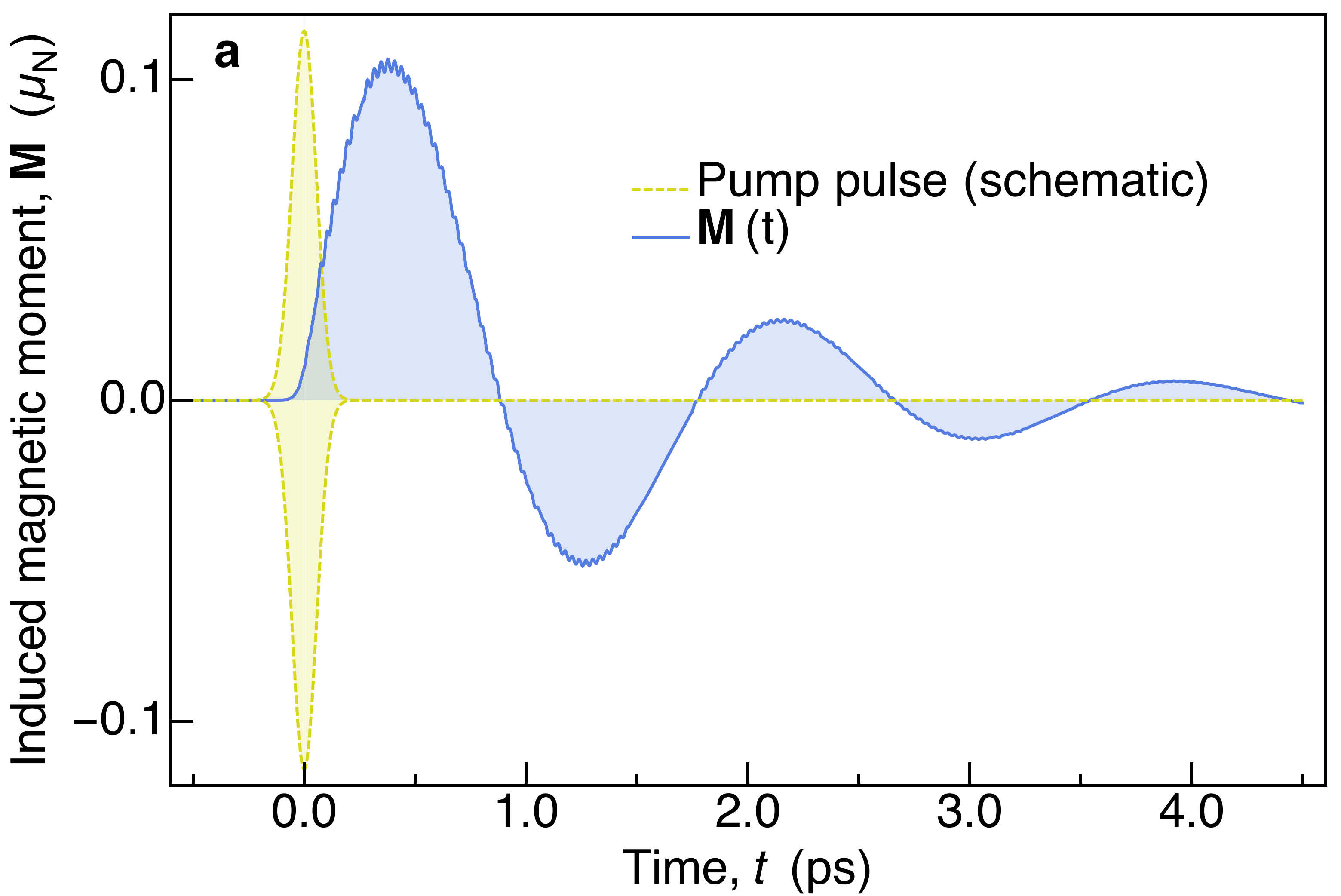}\hspace{10pt}
\includegraphics[scale=0.27]{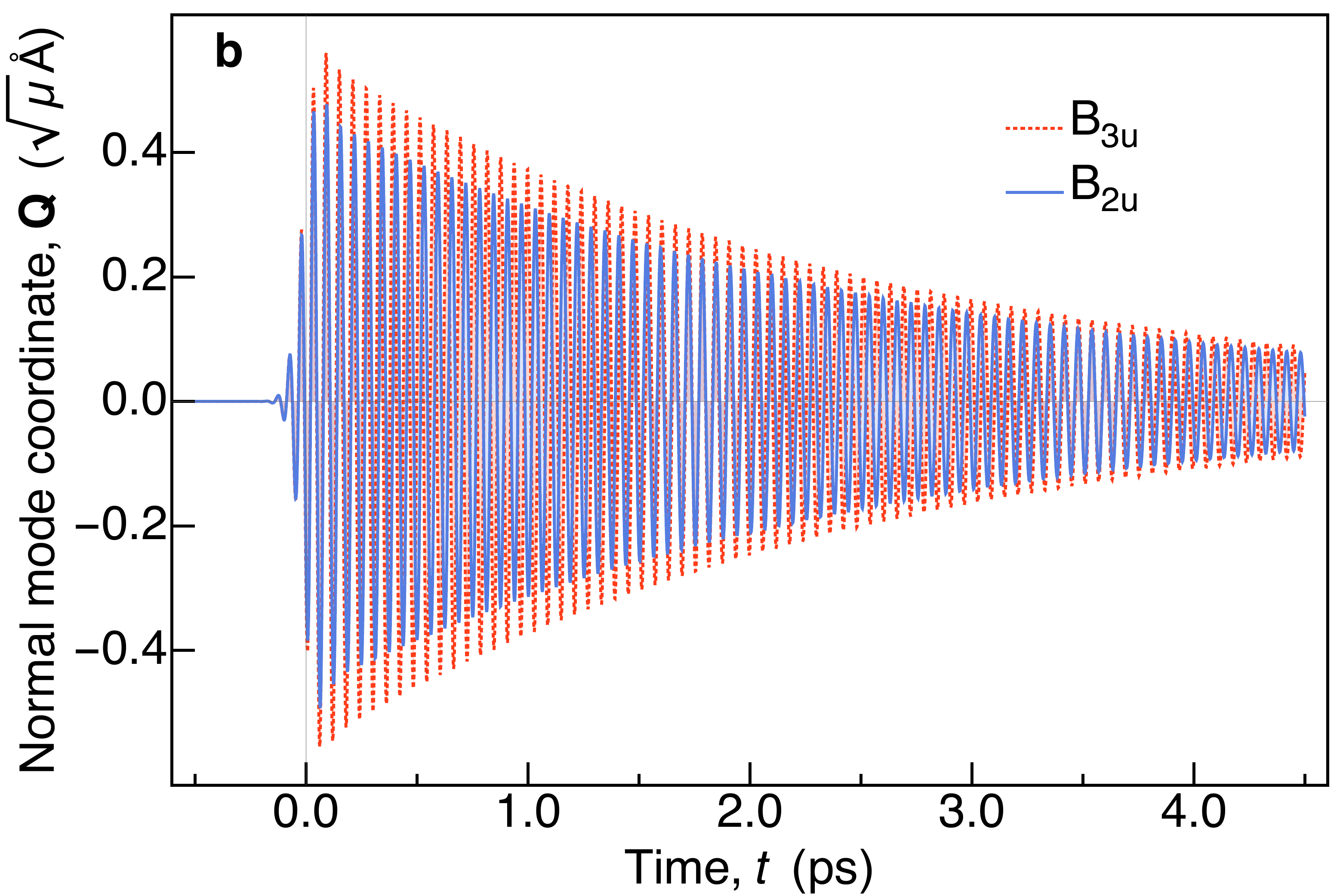}
\caption{
\label{fig:timeevolution}
\textbf{Calculated time-varying magnetic moment from optically driven phonons in ErFeO$_3$.}
\textbf{(a)} Time evolution of the phonon magnetic moment $\mathbf{M}$ in units of the nuclear magneton 
$\mu_{\text{N}}$ in ErFeO$_3$ after excitation with an ultrashort terahertz pulse. The phonon magnetic moment oscillates 
with a large amplitude with the difference frequency, $\omega_{-}=\omega_{1}-\omega_{2}$, and with a small amplitude 
with the sum frequency, $\omega_{+}=\omega_{1}+\omega_{2}$. The schematic pump pulse has a duration of \units{130}{fs}.
\textbf{(b)} Time evolution of the normal mode coordinates $Q_{\alpha}$ of the excited high frequency IR modes 
($\alpha=\text{B}_{3\text{u}},~\text{B}_{2\text{u}}$) in units of $\sqrt{\mu}\ang$, where $\mu$ is the atomic mass 
unit.
}
\end{figure*}


\subsection*{Resonant magnon excitation by\\ optically driven phonons}

A recently observed example of the dynamical multiferroic effect is the resonant excitation of magnons through 
the frequency-dependent magnetic moment of a system of phonons. Consider two nondegenerate phonon modes 
$(\omega_1\neq\omega_2)$ with perpendicular polarity. Setting $\varphi=0$ without loss of generality, 
Eq.~(\ref{eq:magneticmoment}) then reduces to
\begin{eqnarray}\label{eq:magmom_magnon}
\mathbf{M}(t)  \sim  \bigg[ \frac{\omega_{+}}{2} \sin(\omega_{-}t) - \frac{\omega_{-}}{2}\sin(\omega_{+}t) \bigg] 
A_1 A_2 \hat{z}.
\end{eqnarray}
We obtain an induced magnetization that varies in time as a superposition of a large-amplitude oscillation with 
the difference frequency, $\omega_{-}$, and a small-amplitude oscillation with the sum frequency, $\omega_{+}$.

Such a situation was recently realized experimentally by Nova \textit{et al.} by applying an intense linearly 
polarized terahertz pulse along the [110] direction in the $a$-$b$ plane of orthorhombic perovskite-structure erbium 
ferrite, ErFeO$_3$ \cite{nova:2017}. The pulse drives the high-frequency IR-active phonon modes B$_{3\text{u}}$ and 
B$_{2\text{u}}$ with polarization along the $a$ and $b$ lattice vectors. Due to the $Pbnm$ symmetry of ErFeO$_3$, 
these modes have slightly different eigenfrequencies (17.0 and \units{16.2}{THz}). Intriguingly, the spontaneous 
excitation of a magnon at \units{0.75}{THz}, close to the difference frequency, was observed. Our analysis above 
indicates that this behavior is a manifestation of the dynamical multiferroic effect: The large-amplitude part of 
the induced magnetic moment, $\mathbf{M}$, oscillates with the difference frequency $\omega_{-}=\units{0.8}{THz}$, 
sufficiently close to the \units{0.75}{THz} magnon of ErFeO$_3$ \cite{koshizuka:1980}, which is thus resonantly 
excited by $\mathbf{M}$. 

Indeed a dynamical simulation of the evolution of the phonons with parameters computed from density functional theory 
confirms the behaviour predicted phenomenologically in Eq.~(\ref{eq:magmom_magnon}). (See the ``Computational 
methods for phonon calculations'' section for details of the calculation.) In Fig.~\ref{fig:timeevolution}~\textbf{a} 
we show the induced magnetic moment as calculated by Eq.~(\ref{eq:magmom_phononic}) and in 
Fig.~\ref{fig:timeevolution}~\textbf{b} the normal mode coordinate of the 
high-frequency IR-active B$_{3\text{u}}$ and B$_{2\text{u}}$ modes following an excitation with an ultrashort terahertz 
pulse. The small-amplitude oscillation of \textbf{M} with the sum frequency is negligible and only the large-amplitude 
oscillation with the difference frequency contributes. The induced magnetic moment peaks at the order of magnitude of 
the nuclear magneton with $|\mathbf{M}|\approx 0.1\mu_{\text{N}}$ per unit cell. It decays together with the IR modes 
over the timescale of a few picoseconds; experimentally the induced magnon was found to survive for at least 
\units{35}{ps} \cite{nova:2017}.

While the phonon Zeeman effect is independent of the magnitude of the phonon magnetic moment, the resonant effect 
discussed here is quadratic in the amplitude of the dynamical polarization and therefore dependent on the intensity
of the exciting terahertz pulse. At high pulse intensities an anharmonic coupling of the excited IR phonons to 
Raman-active phonons was shown to be relevant in ErFeO$_3$ \cite{juraschek:2017}. These modes do not contribute to the
gyromagnetic ratio however, and we can therefore neglect the effect of nonlinear phononics in the above analysis.


\subsection*{Computational methods for phonon calculations}

We calculated the phonon eigenfrequencies and eigenvectors, and the Born effective charges of SrTiO$_3$ and ErFeO$_3$ 
from first-principles using the density functional theory formalism as implemented in the Vienna ab-initio simulation 
package (VASP) \cite{kresse:1996,kresse2:1996} and the frozen-phonon method as implemented in the phonopy package 
\cite{phonopy}. We used the default VASP PAW pseudopotentials with Er 4$f$ electrons treated as core states. We 
converged the Hellmann-Feynman forces to \units{10^{-5}}{eV/\ang{}} using a plane-wave energy cut-off of 
\units{700}{eV} and a 7$\times$7$\times$5 $k$-point mesh to sample the Brillouin zone for SrTiO$_3$ and \units{850}{eV}, 
6$\times$6$\times$4 for ErFeO$_3$. For the exchange-correlation functional we chose the PBEsol \cite{PBEsol} form 
of the generalized gradient approximation (GGA) and imposed a Hubbard correction 
of $U=\units{3.7}{eV}$ and a Hund's exchange of $J=\units{0.7}{eV}$ on the Fe $3d$ states in ErFeO$_3$. Our fully 
relaxed structures with lattice constants $a=\units{5.51}{\ang}$ and $c=\units{7.77}{\ang}$ for SrTiO$_3$ and 
$a=\units{5.19}{\ang}$, $b=\units{5.56}{\ang}$, and $c=\units{7.52}{\ang}$ for ErFeO$_3$ fit reasonably well to the 
experimental values of Refs.~\cite{eibschutz:1965,kiat:1996}, as do our calculated phonon eigenfrequencies 
\cite{fleury:1968,galzerani:1982,subbarao:1970,koshizuka:1980,nova:2017}. Our calculated values for the highest IR 
phonon frequencies are \units{16.52}{THz} for the B$_{3\text{u}}$ and \units{15.95}{THz} for the B$_{2\text{u}}$ mode 
in ErFeO$_3$. For the time evolution of IR phonons after a pulsed optical excitation, we obtain the time-dependent 
normal mode coordinates, $Q$, by numerically solving the dynamical equations of motion:
\begin{equation}\label{eq:EOM}
\ddot{Q}_{\alpha} + \kappa_{\alpha} \dot{Q}_{\alpha} + \omega^{2}_{\alpha} Q_{\alpha} = F,
\end{equation}
where $\kappa_{\alpha}$ is the friction coefficient and $\omega_{\alpha}$ the eigenfrequency of mode $\alpha$. 
The periodic driving force $F$ models the terahertz pulse in a realistic fashion with gaussian shape with an 
amplitude of 10\,MV\,cm$^{-1}$, and a finite width in both time (fwhm=\units{130}{fs}) and frequency (peak 
frequency \units{19.5}{THz}, fwhm=\units{6.5}{THz}) \cite{nova:2017}.


\section*{Beyond lattice dynamics}

In the previous two examples we showed how the magnetic moment arising from a system of phonons couples lattice and 
magnetic degrees of freedom. In the following we discuss examples in which the time-dependent polarization is not 
caused by lattice dynamics.

\subsection*{Dzyaloshinskii-Moriya-type electromagnons}

The existence of electromagnons -- that is spin waves excited by a.c. electric fields -- was demonstrated ten 
years ago in multiferroic materials with an incommensurate (cycloidal or helicoidal) magnetic structure 
\cite{pimenov:2006}. In the original report, the interaction was shown to be mediated through conventional 
Heisenberg coupling of spins, leading to a phonon-magnon hybridization that makes the magnon electroactive (and 
is therefore called an electromagnon) 
\cite{pimenov:2006,senff:2007,sushkov:2008,kida:2008,valdesaguilar:2009,pimenov:2009,lee:2009,takahashi:2012}. 
A second mechanism for generating electromagnons, in which electroactive excitation of the spin spiral occurs 
via the Dzyaloshinskii-Moriya interaction, has also been identified 
\cite{katsura:2007,pimenov:2008,shuvaev:2010,takahashi:2012,takahashi:2013,shuvaev:2013}. We show in the following 
that the description of these electromagnons from the viewpoint of dynamical multiferroicity is equivalent to 
the previously introduced formalism via the inverse effect of the Dzyaloshinskii-Moriya interaction \cite{katsura:2007}.

\begin{figure}[t]
\includegraphics[scale=0.25]{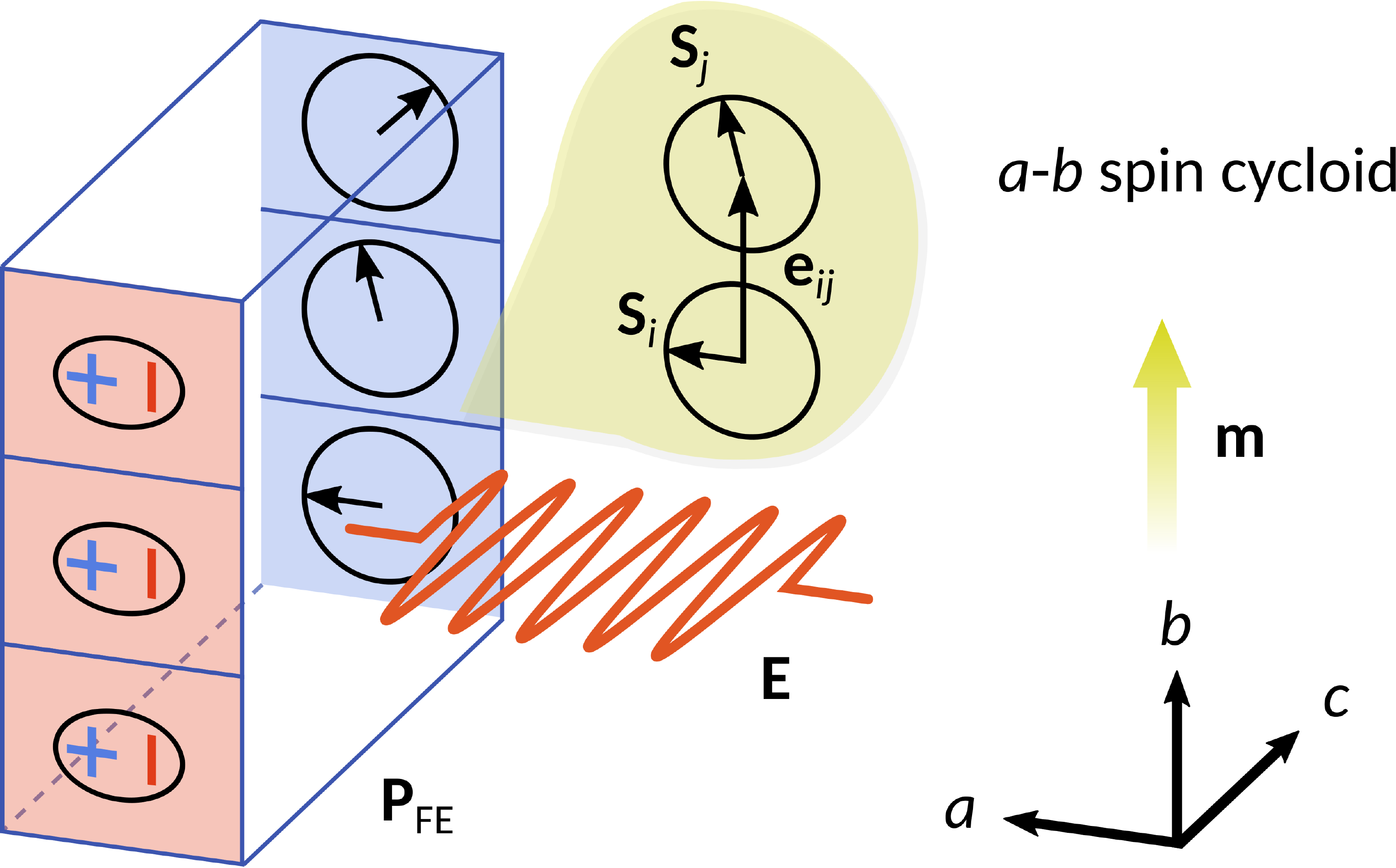}
\caption{
\label{fig:abcycloid}
\textbf{Dzyaloshinskii-Moriya-type electromagnon symmetry in TbMnO$_3$.} 
Spin cycloid oriented in the $a$-$b$ plane with ferroelectric polarization $\mathbf{P}_\text{FE}$ along $a$. To 
excite a Dzyaloshinskii-Moriya electromagnon, the a.c. electric field of the sub-terahertz radiation \textbf{E} has 
to be aligned along $c$ and the induced magnetization \textbf{m} is along $b$.
}
\end{figure}

The standard description of a Dzyaloshinskii-Moriya-type electromagnon in a helical magnet is through the coupling of 
the spin degrees of freedom of the cycloid, $\mathbf{S}_m$, with a uniform lattice displacement, $\mathbf{u}$. The 
ferroelectric polarization is given by 
$\mathbf{P}_\text{FE}\propto\mathbf{e}_{ij}\times(\mathbf{S}_i\times\mathbf{S}_j)$, where $\mathbf{e}_{ij}$ is 
the vector connecting the sites $i$ and $j$. The component $n$ of the lattice displacement that couples to the 
rotation of the spin plane is parallel to its rotation axis, $\mathbf{u}^n~||~(\mathbf{S}_i\times\mathbf{S}_{j})$. 
An a.c. electric field, $\mathbf{E}~||~\mathbf{u}^n$, then induces a magnetization, $\mathbf{m}~||~\mathbf{e}_{ij}$, 
and excites an electromagnon when it matches the frequency of the eigenmode of the spin cycloid \cite{katsura:2007}. 
Specifically in TbMnO$_3$ with a magnetic field applied along $b$, the spin cycloid lies in the $a$-$b$ plane. Since 
$(\mathbf{S}_i\times\mathbf{S}_j)~||~c$ and $\mathbf{e}_{ij}~||~b$, the resulting $\mathbf{P}_\text{FE}~||~a$. The a.c. 
electric field component of sub-terahertz radiation, $\mathbf{E}~||~c$, then induces a magnetization, $\mathbf{m}~||~b$, 
and excites a Dzyaloshinskii-Moriya-type electromagnon when it matches the eigenfrequency of the spin cycloid at 
\units{0.63}{THz} \cite{shuvaev:2010}. To disentangle this clutter of alignments, we illustrate the symmetry in 
Fig.~\ref{fig:abcycloid}. Note that a cooperative contribution of the ferroelectric polarization from symmetric 
magnetostriction has been reported \cite{mochizuki:2009,mochizuki2:2010}, which, however, leaves our symmetry 
analysis here unaffected.

This situation is exactly consistent with the formalism of dynamical multiferroicity when the two perpendicular 
polarizations in Eq.~(\ref{eq:polarization}) are given by $\mathbf{P}=(P_1(0),0,P_2(t))$, where 
$P_1(0)\equiv\mathbf{P}_\text{FE}$ ($\omega_1=0$, $\varphi=\pi/2$) and $P_2(t)\equiv\mathbf{E}(t)$ 
($\omega_2=\omega_0$). Eq.~(\ref{eq:magneticmoment}) then reduces to 
\begin{eqnarray}\label{eq:magmom_electromagnon}
\mathbf{M}(t)  \sim  \omega_0 \cos(\omega_0 t) A_1 A_2 \hat{y}.
\end{eqnarray}
The formalism predicts an induced magnetization that oscillates with the frequency of the terahertz radiation, $\omega_0$, 
identical to the magnetization described above within the inverse Dzyaloshinskii-Moriya formalism, 
$\mathbf{M}(t)\equiv\mathbf{m}$.

\begin{table*}[t]
\centering
\bgroup
\def\arraystretch{1.3}
\caption{
\textbf{Summary of examples for the dynamical multiferroic effect.} The table shows the frequency of the induced 
magnetic moment, \textbf{M}$(t)$, the type of excitation, the type of drive, and the order of the effect in terms 
of the time-dependent polarization, $P(t)$, for the four examples discussed in this work. From left to right: Phonon 
Zeeman effect in SrTiO$_3$, Resonant magnon excitation by optically driven phonons in ErFeO$_3$, 
Dzyaloshinskii-Moriya-type electromagnon in TbMnO$_3$, and the inverse Faraday effect in DyFeO$_3$.
}
\begin{tabular}{c K{3.5cm} K{3.5cm} K{3.5cm} K{3.5cm}}
\hline\hline
  & 
\begin{minipage}{\linewidth}
\includegraphics[scale=0.25]{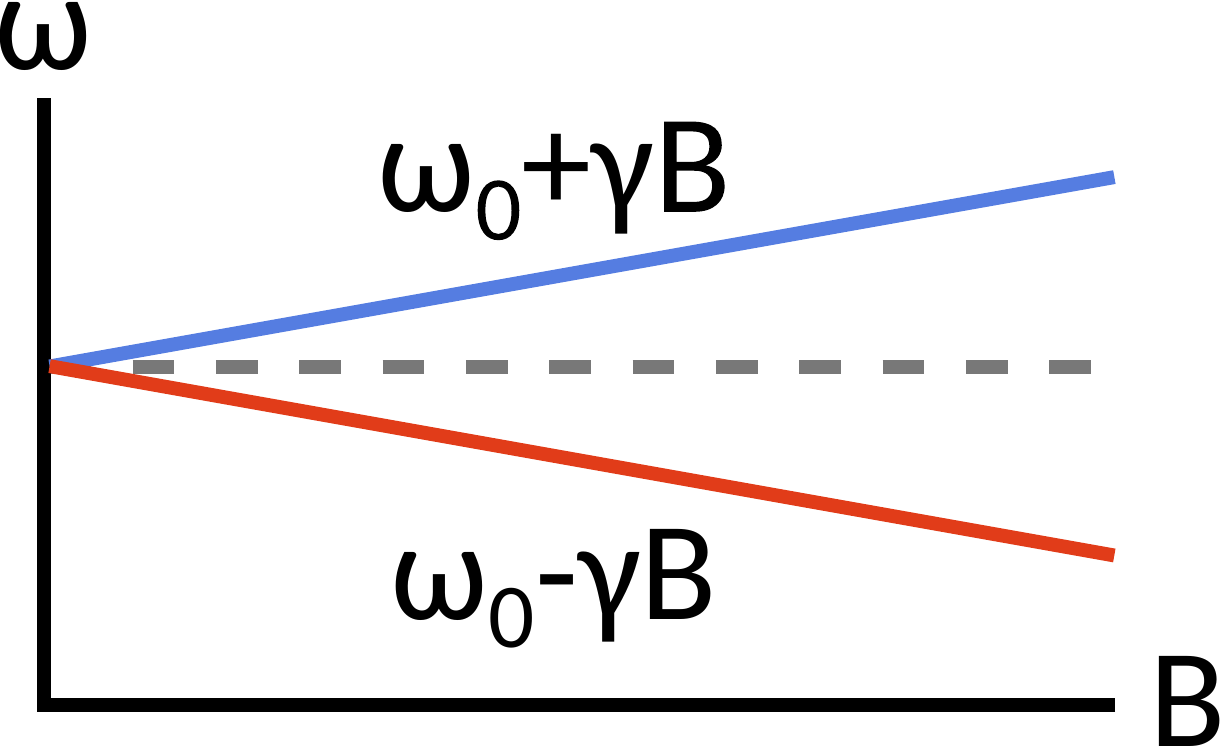}
\end{minipage} &
\begin{minipage}{\linewidth}
\includegraphics[scale=0.25]{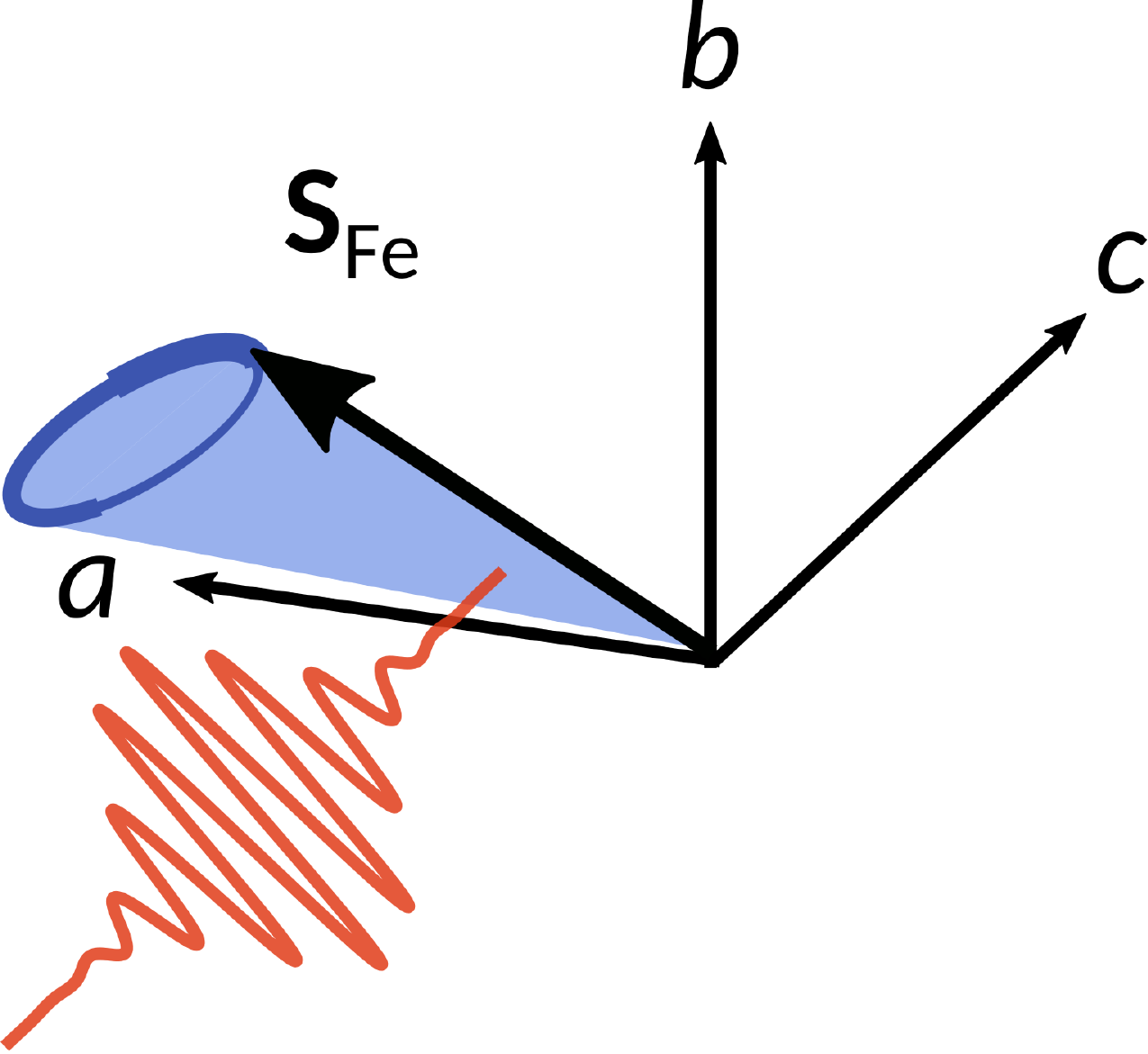}
\end{minipage} &
\begin{minipage}{\linewidth}
\includegraphics[scale=0.25]{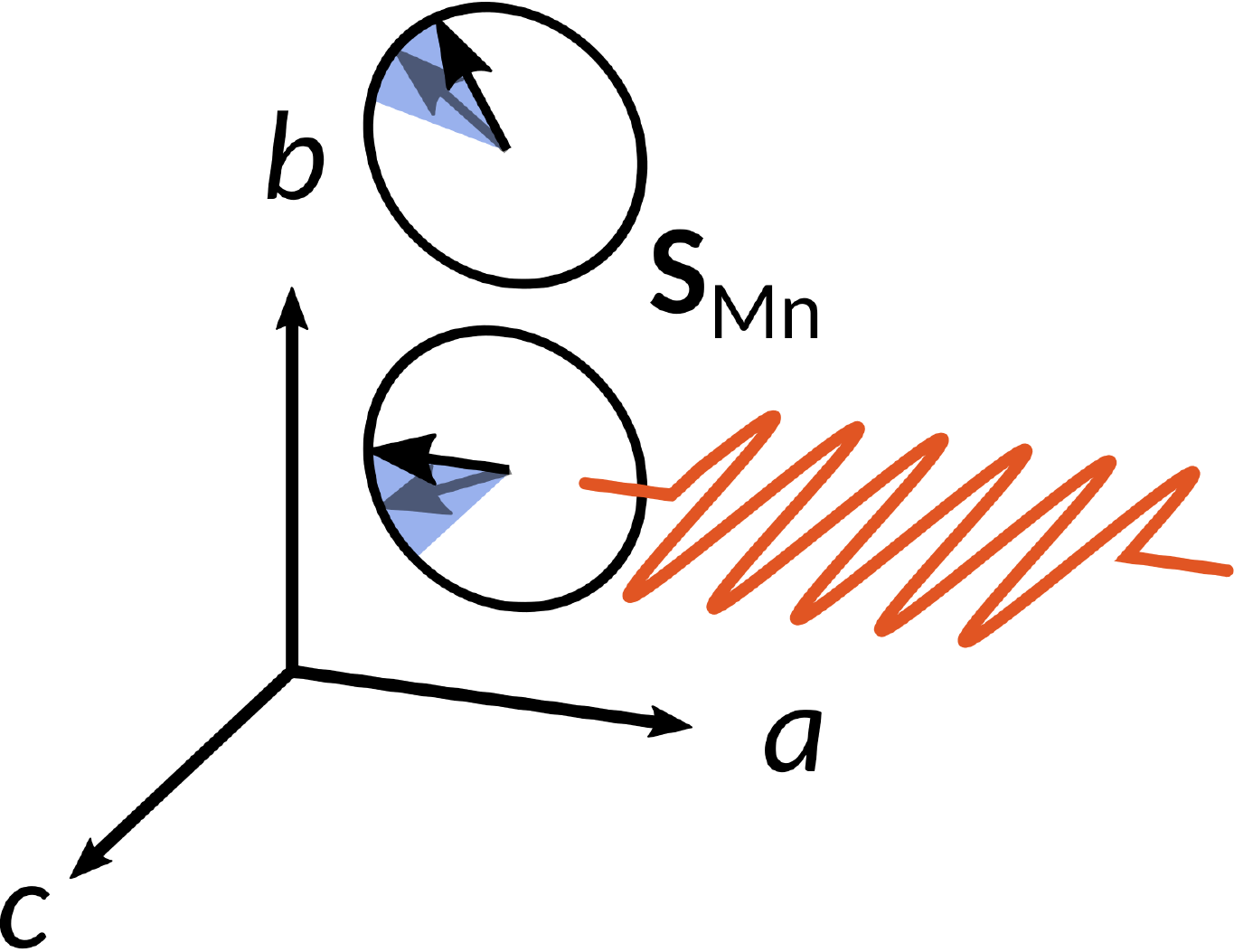}
\end{minipage} &
\begin{minipage}{\linewidth}
\includegraphics[scale=0.25]{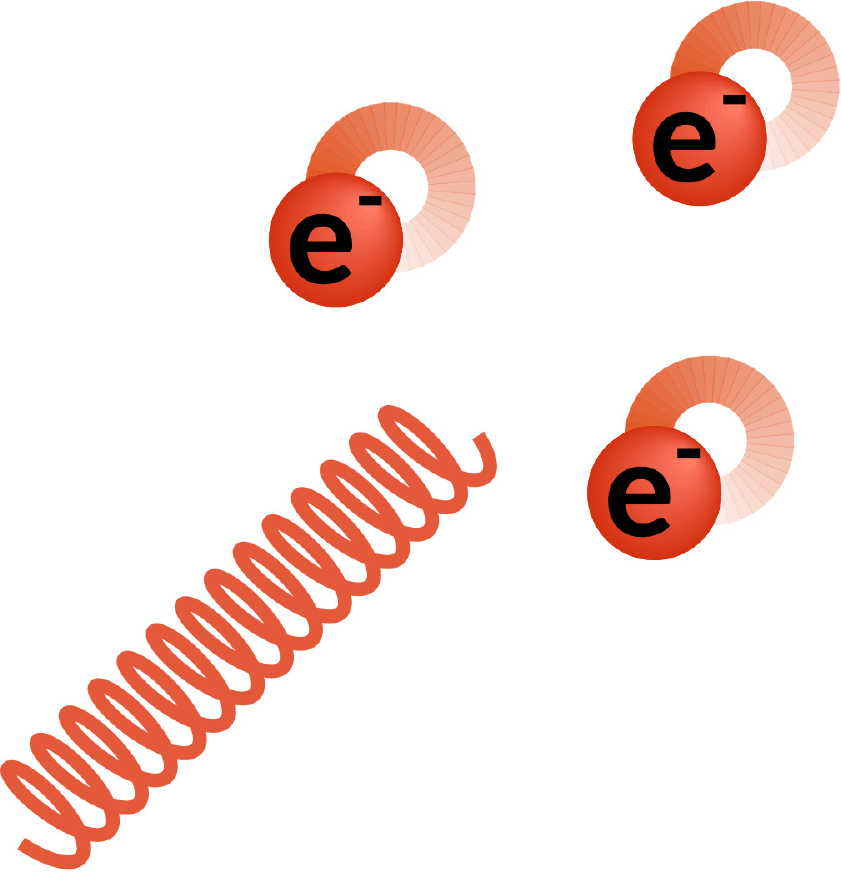}
\end{minipage} \\ \hline
Example material & SrTiO$_3$ & ErFeO$_3$ & TbMnO$_3$ & DyFeO$_3$ \\
$\mathbf{M}(t)$ frequency & static & $\omega_1-\omega_2$ & $\omega_0$ & static  \\
Excitation & thermal, zero-point & THz & sub-THz & IR \\
Polarization of drive & -- & linear & linear & circular\\
Order of P(t) & -- & quadratic & linear & quadratic\\
\hline\hline
\end{tabular}
\label{tab:orderofexcitations}
\egroup
\end{table*}


\subsection*{Inverse Faraday effect}

Finally, we show that the inverse Faraday effect, which describes the generation of a static magnetization 
in a material when irradiated with circularly polarized light, can also be interpreted from the viewpoint of 
dynamical multiferroicity. The inverse Faraday effect was first demonstrated in the 1960's 
\cite{pitaevskii:1961,vanderziel:1965,pershan:1966} and experienced a massive revival in the last decade, after 
it was shown that it can be used to control magnetization non-thermally with photomagnetic pulses \cite{kimel:2005}. 
The standard description of the effect is 
\begin{equation}\label{eq:IFE}
\mathbf{M}(\omega) = \chi(\omega) \mathbf{E}(\omega) \times \mathbf{E^\ast}(\omega),
\end{equation}
where $\mathbf{M}$ is the induced magnetization, $\chi$ the magneto-optic susceptibility and $\mathbf{E}$ the 
electric field component of the circularly polarized light. The electric field induces perpendicular time-varying 
polarizations in the material and the connection to dynamical multiferroicity is revealed by rewriting 
Eq.~(\ref{eq:IFE}) in the time domain:
\begin{eqnarray}\label{eq:IFE_DynMult}
\mathbf{M}(t) & = & \chi(t) \mathbf{E}(t) \times \partial_{t}\mathbf{E}(t) \nonumber\\
              & \sim & \mathbf{P} \times \partial_{t}\mathbf{P}.
\end{eqnarray}
The perpendicular time-dependent polarizations induced by the circularly polarized light both oscillate with the 
frequency of the light $(\omega_1=\omega_2=\omega_0)$ and are shifted in phase by $\varphi=\pi/2$. 
Eq.~(\ref{eq:magneticmoment}) therefore reproduces the induced static magnetization of the inverse Faraday effect:
\begin{eqnarray}\label{eq:magmom_IFE}
\mathbf{M}(0) \sim \omega_0 A_1 A_2 \hat{z}.
\end{eqnarray}
Note that the investigation of the inverse Faraday effect has been extended in recent years by microscopic theories 
\cite{battiato:2014} that also describe ultrafast timescales \cite{reid:2010,popova:2011}, as well as the case of 
linearly polarized light \cite{ali:2010}; these extensions are beyond the scope of the analogy provided here.


\section*{Discussion}

In summary, we have introduced the concept of dynamical multiferroicity and shown that it provides a straightforward 
unifying interpretation of four diverse phenomena, summarized in Table ~\ref{tab:orderofexcitations}: i) the Zeeman 
splitting of phonon spectra in a magnetic field, ii) resonant magnon scattering, iii) electromagnon coupling, and iv) 
the inverse faraday effect. In the case of i) and ii), in which the multiferroicity is mediated by lattice dynamics, 
we showed that the dynamical multiferroic coupling can be calculated quantitatively using standard density functional 
theory methods, and we performed such calculations for representative materials in each case. 

Since the term \textit{nonlinear} is used in both optics (for example second-harmonic generation) and phononics (for 
example cubic and higher-order phonon-phonon interactions), we also clarify the order (linear or higher order and in 
which parameter) of the excitations for the four cases. First, in the cases of SrTiO$_3$, ErFeO$_3$ and DyFeO$_3$, the 
materials are centrosymmetric, so a nonlinear second-harmonic optical response to the terahertz pulse is not possible. 
In the case of TbMnO$_3$, second-harmonic generation occurs at higher energies than the sub-THz radiation used in the 
example (see for example Ref.~\cite{matsubara:2015}). In addition, while cubic and higher-order non-linear phononic 
coupling has indeed been demonstrated in some of these materials (see for example, Ref.~\cite{juraschek:2017} for a 
detailed analysis in the case of ErFeO$_3$), the phenomena described here are not driven by non-linear phononic effects. 
The phonon Zeeman effect in SrTiO$_3$ is independent of the amplitude of the degenerate phonons and requires no external 
excitation and so is zeroth-order in the polarization amplitude. The magnon excitation in ErFeO$_3$ depends on the 
quadratic term in the phonon amplitude and so is nonlinear in the amplitude of the driven time-dependent polarization, 
$P(t)$. As for the Dzyaloshinskii-Moriya-type electromagnon excitation, the induced magnetization is \textit{linear} in 
$P(t)$ (with the perpendicular ferroelectric polarization being static), while the inverse Faraday effect is nonlinear 
in $P(t)$ due to the excitation with circularly polarized light.

The phonon Zeeman splitting that we analyze here is distinct from previously discussed interactions 
between phonons and magnetism in \textit{magnetic} materials. These include the phonon angular momentum arising from 
spin-phonon interaction \cite{zhang:2014}, the splitting of acoustic phonons due to the spin-orbit interaction 
\cite{liu:2017}, the splitting of nondegenerate Raman-active phonon modes accompanying a ferroelectric to paraelectric 
phase transition \cite{rovillain:2011} and the magnetic-field dependence of the phonon frequencies in Ce compounds 
\cite{schaack:1975,schaack:1976,schaack:1977}. In contrast, the phonon Zeeman effect does not require magnetic ions. 
The report of a splitting of degenerate phonons due to the interaction with magnetoexcitons in graphene \cite{remi:2014} 
is a fascinating phenomenon, but also a different mechanism from the phonon Zeeman effect. It is further distinct from 
the phonon Hall effect proposed for acoustic phonons as an analogue of the anomalous Hall effect 
\cite{strohm:2005,sheng:2006}.

With the increased availability of intense THz sources of radiation, we anticipate that additional manifestations of 
dynamical multiferroicity will be revealed over the next years. Reciprocally, we expect that the effect will be used 
to engineer new behaviors that are not accessible in the static domain. In this context, we point out that dynamical 
multiferroicity provides a unit-cell analogue of an electric motor, in which the time-dependent polarization acts as 
a nanoscale electromagnetic coil to generate magnetic fields. This analogy might open a pathway to unforeseen 
technological applications.


\begin{acknowledgments}
We thank G. Aeppli, A. Cavalleri,  M. Fiebig, T. F. Nova, and  A. Scaramucci for useful discussions. This work was 
supported by the ETH Z\"urich, by Dr. Max R\"ossler and the Walter Haefner Foundation through the ETH Z\"urich
Foundation, by US DoE E3B7 and by the ERC Advanced Grant program numbers 291151 and DM-321031. Calculations were performed at the 
Swiss National Supercomputing Centre (CSCS) supported by the project IDs s624 and p504.
\end{acknowledgments}


\bibliography{Nicola.bib}

\end{document}